\documentclass[%
reprint,
superscriptaddress,
amsmath,amssymb,
aps,
]{revtex4-2}

\usepackage{graphicx}

\usepackage{verbatim} 
\usepackage{dcolumn}
\usepackage{siunitx}
\usepackage{multirow} 

\begin{document}
	
    \preprint{}

    \title{Viscoelastic materials are most energy efficient when loaded and unloaded at equal rates}

    \author{Lucien Tsai}
    \affiliation{Department of Physics, Harvey Mudd College, Claremont, CA, USA, 91711}
    \author{Paco Navarro}
    \affiliation{Department of Physics, Harvey Mudd College, Claremont, CA, USA, 91711}
    \author{Siqi Wu}
    \affiliation{Department of Physics, Harvey Mudd College, Claremont, CA, USA, 91711}
    \author{Taylor Levinson}
    \affiliation{Department of Physics, Harvey Mudd College, Claremont, CA, USA, 91711}
    \author{Elizabeth Mendoza}
    \affiliation{Department of Ecology and Evolutionary Biology, University of California, Irvine, CA, USA, 92697}
    \author{M. Janneke Schwaner}
    \affiliation{Department of Ecology and Evolutionary Biology, University of California, Irvine, CA, USA, 92697}
    \author{Monica A. Daley}
    \affiliation{Department of Ecology and Evolutionary Biology, University of California, Irvine, CA, USA, 92697}
    \author{Emanuel Azizi}
    \affiliation{Department of Ecology and Evolutionary Biology, University of California, Irvine, CA, USA, 92697}
    \author{Mark Ilton}
    \email{milton@g.hmc.edu}
    \affiliation{Department of Physics, Harvey Mudd College, Claremont, CA, USA, 91711}
    
    \date{\today}

\begin{abstract}
 Biological springs can be used in nature for energy conservation and ultra-fast motion. The loading and unloading rates of elastic materials can play an important role in determining how the properties of these springs affect movements. We investigate the mechanical energy efficiency of biological springs (American bullfrog plantaris tendons and guinea fowl lateral gastrocnemius tendons) and synthetic elastomers. We measure these materials under symmetric rates (equal loading and unloading durations) and asymmetric rates (unequal loading and unloading durations) using novel dynamic mechanical analysis measurements. We find that mechanical efficiency is highest at symmetric rates and significantly decreases with a larger degree of asymmetry. A generalized 1D Maxwell model with no fitting parameters captures the experimental results based on the independently-characterized linear viscoelastic properties of the materials. The model further shows that a broader viscoelastic relaxation spectrum enhances the effect of rate-asymmetry on efficiency. Overall, our study provides valuable insights into the interplay between material properties and unloading dynamics in both biological and synthetic elastic systems.
\end{abstract}
	

\maketitle

\begin{figure*}[tb]
\centering
\includegraphics[width=\linewidth]{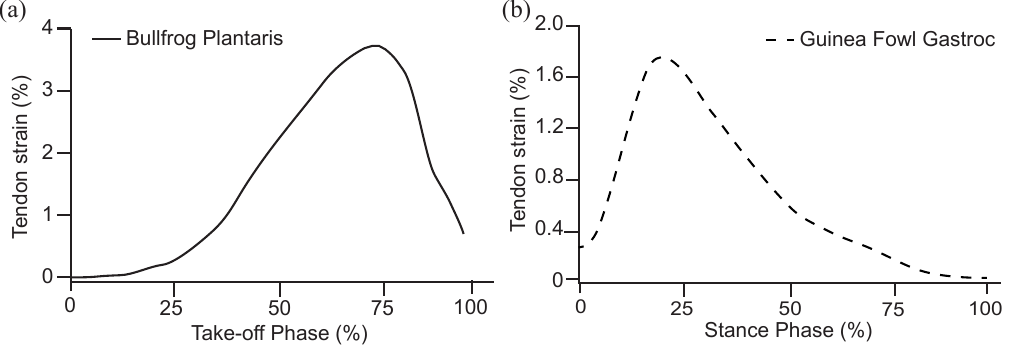}
\caption{\label{fig0} \textbf{Strains patterns experienced by tendons in vivo during latch-mediated spring actuation (a) and cyclic locomotion (b).} \textbf{(a)} Tendon strains in the plantaris (gastrocnemius) tendon of bullfrogs plotted against the proportion of the take-off phase of a single representative jump. The maximum strain rates during the unloading phase is about 3.5 times faster than the loading phase. \textbf{(b)} Strain of the lateral gastrocnemius tendon of guinea fowl during the stance phase of a representative running trial. The maximum tendon strain rates experienced during the loading are faster that that those observed during the unloading phase. These plots highlight the variation in observed strain patterns during different modes of locomotion. Data for panels (a) and (b) are recreated from refs.~\cite{mendoza2023mechanical} and \cite{schwaner2023linking} respectively. }
\end{figure*}

Biological springs perform a variety of functions in natural movements including energy conservation and latch-mediated spring actuation~\cite{robertsFlexibleMechanismsDiverse2011,highamSpringsSteroidsSlingshots2013,iltonPrinciplesCascadingPower2018,longoPowerAmplificationLatchmediated2019}. Energy-conserving movements involve cyclic, repeating locomotor patterns such as terrestrial running or hopping gaits~\cite{alexanderMechanicsHoppingKangaroos1975,cavagnaMechanicalWorkTerrestrial1977,heglundEnergeticsMechanicsTerrestrial1982}, insect flight~\cite{dickinsonMuscleEfficiencyElastic1995,youngDetailsInsectWing2009}, and fish swimming~\cite{fishPassiveActiveFlow2006}. In contrast, latch-mediated spring actuation is characterized by a rapid release of energy such as the movements of jumping frogs~\cite{robertsWeakLinkMuscle2011,astleyEvidenceVertebrateCatapult2012} and insects~\cite{burrowsMorphologyActionHind2006,suttonBiomechanicsJumpingFlea2011,burrowsLocustsUseComposite2012,bolminLatchingClickBeetle2019}, raptorial appendage strikes of mantis shrimps~\cite{patekDeadlyStrikeMechanism2004,patekComparativeSpringMechanics2013}, tongue projections of chameleons~\cite{degrootEvidenceElasticProjection2004,andersonScalingBallisticTongue2012,andersonShotScalingBallistic2016}, and mandible closures of trap-jaw ants~\cite{patekMultifunctionalityMechanicalOrigins2006}. These diverse functions often result in variations in the relative rate of loading and unloading of elastic structures during a movement. Elastic structures used for latch-mediated spring actuation are often loaded at slow rates compared to the rapid unloading rates associated with elastic recoil (Fig.~\ref{fig0}a). In contrast, the behavior of elastic structures during cyclic movements are more varied. Previous work has shown symmetrical patterns with near equal strain rates observed during loading and unloading ~\cite{harrison2010relationship}. Others have shown that tendon strain rates are highest during loading, at the beginning of the stance phase (Fig.~\ref{fig0}b). The diverse patterns of loading and unloading rates observed in biological systems suggest that shifts in mechanical properties arising from rate-asymmetry are likely to have significant performance consequences.

For vertebrates, tendon is a tissue often characterized as a biological spring that operates in series with skeletal muscle~\cite{robertsFlexibleMechanismsDiverse2011} and exhibits a highly nonlinear elastic response. When stretched, tendon is relatively compliant at low strain and stiffens before reaching a roughly linear elastic regime at approximately 3-10\% strain~\cite{gosline2018mechanical}. This strain-stiffening response enables significant elastic energy storage in tendon to drive latch-mediated spring actuated movements~\cite{robertsWeakLinkMuscle2011,astleyEvidenceVertebrateCatapult2012,mendoza2021tuned}. Tendons also operate with relatively high mechanical energy efficiency which allows them to conserve energy during cyclical movements~\cite{robertsFlexibleMechanismsDiverse2011,harrison2010relationship}.
	
Previous work on the mechanical energy efficiency of tendons has focused on symmetric loading and unloading rates. The primary metric used to describe a material's energy efficiency is its resilience, defined as the ratio of the energy released divided by the energy stored during a single loading and unloading cycle. Tendon has been shown to have a 90-95\% resilience over a range of symmetric rates ($\sim$ \SI{0.1}{Hz} to \SI{10}{Hz})~\cite{kerDynamicTensileProperties1981,matsonTendonMaterialProperties2012,rosario2020loading} with a range of resilience from 80\% to 97\% measured across a variety of tendons~\cite{pollockRelationshipBodyMass1994}. However, resilience can depend on how it is measured. For example, when measured as a mass-spring system, tendons from the feet of sheep have a significantly lower resilience (62\% at \SI{5}{Hz})~\cite{cumingReboundResilienceTendons1978}.

Although the literature has historically focused on tendon stretched at symmetric rates, the growing body of literature suggests that tendons undergo diverse patterns of asymmetric loading rates during natural behaviors (Fig.~\ref{fig0}). These observed differences in loading and unloading rates has motivated studies to explore the response to asymmetric rates in tendon~\cite{rosario2020loading} and other biological springs~\cite{wold2023structural} over a relatively narrow range of asymmetry.  

Here we explore rate-asymmetry over a broad range of timescales to understand its underlying physical principles. In this study we ask: Is there a connection between the biological function of energy-conserving movements and material performance under symmetric rates? How does changing the relative loading and unloading rate affect the mechanical efficiency of tendon? The measurement of materials at asymmetric rates is typically performed using elastic recoil experiments, where a material is slowly loaded followed by a rapid high-rate unloading of elastic waves in the material~\cite{vermorelRubberBandRecoil2007,bogoslovovViscoelasticEffectsFree2007,niemczuraResponseRubbersHigh2011,tunnicliffe2015free,yoon2017application,iltonEffectSizescaleKinematics2019,prado2020achieving,liang2020programming,liang2022dynamic}; however, these measurements require strains larger than those accessible with tendon. 

To address these guiding questions in tendons, we perform novel dynamic mechanical analysis (DMA) measurements using triangular waveforms that either have the same loading and unloading times (symmetric rates) or different loading and unloading times (asymmetric rates) for the same maximum strain. By measuring lateral gastrocnemius tendons from guinea fowl and plantaris tendons from American bullfrogs, we find these tendons are most energy efficient when subjected to symmetric loading and unloading rates. To test the generality of this result and understand its underlying physical cause, we also repeat the same measurements on synthetic elastomer materials with known linear viscoelastic properties and perform simulations. We find the same qualitative result independent of the material and amount of applied deformation: a higher degree of asymmetry between the loading and unloading rates reduces mechanical energy efficiency. Using a 1D linear viscoelastic model, we find that a broad distribution of relaxation times (characteristic of biological materials) enhances the strength of the asymmetric rate effect. Given the generality of our model and results, these findings should be applicable to a wide range of biological materials and could be used to inform the design of efficient elastic mechanisms in soft robotics and bioengineering.

	\begin{figure*}[bt]
		\centering
		\includegraphics[width=\linewidth]{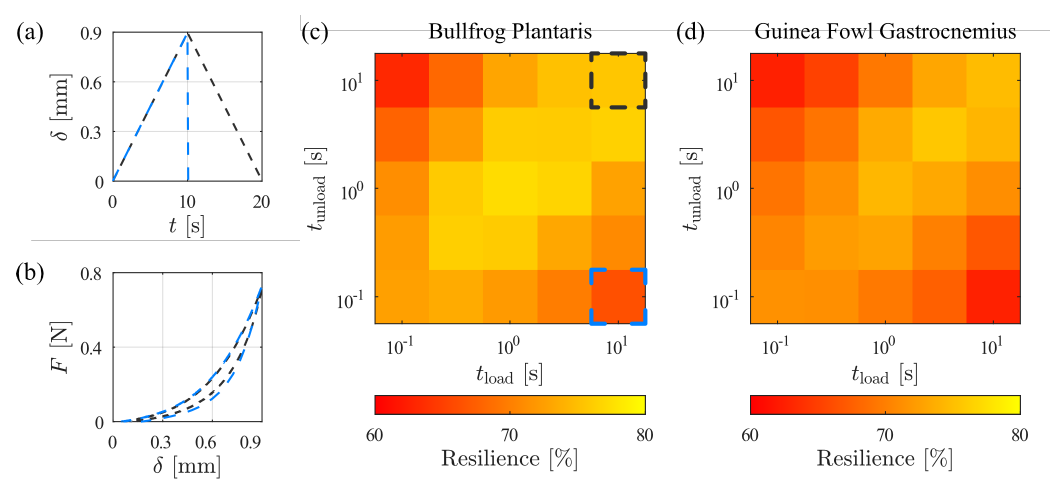}
		\caption{\label{fig1}\textbf{Tendons are most mechanically efficient when loaded and unloaded at equal rates.} \textbf{(a)} The displacement-time ($\delta-t$) profiles for a symmetric rate test where a tendon was loaded for \SI{10}{s} and unloaded for \SI{10}{s} (dashed black curve) and an asymmetric rate test where the tendon was loaded for \SI{10}{s} and unloaded for \SI{0.1}{s} (dashed blue curve). \textbf{(b)} Representative force-displacement ($F-\delta$) responses of bullfrog plantaris tendon to the two displacement-time inputs from panel (a). The tendons exhibit a characteristic strain-stiffening response with a low hysteresis between the loading and unloading curves (high resilience). The tendon has a lower resilience for the asymmetric rate (dashed blue curve) compared to the symmetric rate (dashed black curve). \textbf{(c)} The resilience of bullfrog plantaris tendons were measured for different combinations of loading and unloading rates. The resilience was highest when measured at symmetric rates, which corresponds to the diagonal elements of the heatmap. Each resilience is calculated from an average of 5 tendons. The black and blue dashed squares indicate the rates described in panels (a) and (b). \textbf{(d)} The average resilience of 5 guinea fowl lateral gastrocnemius tendons was consistent with the bullfrog plantaris tendon response. The full data set is available in Supplemental Tables S1 and S2.}
	\end{figure*}

	\section*{Results}

	\begin{figure*}[hbt]
		\centering
		\includegraphics[width=\linewidth]{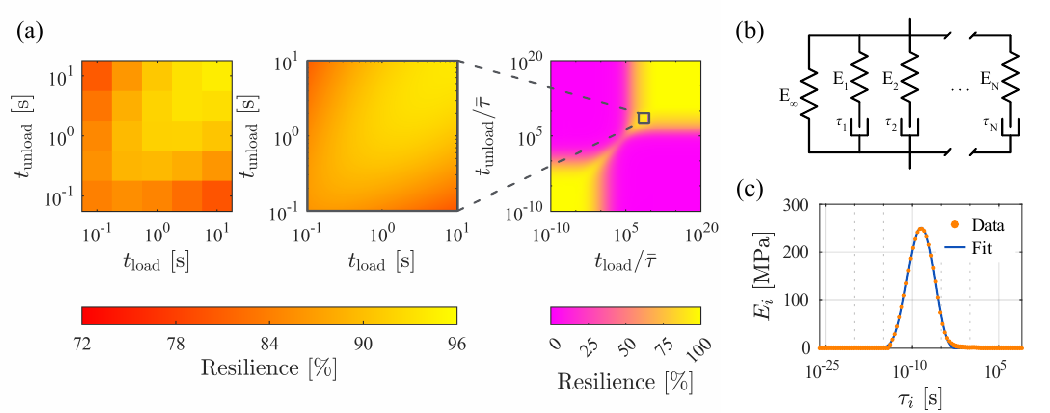}
		\caption{\label{fig2}\textbf{Symmetric loading and unloading rates are also most mechanically efficient for a synthetic polyurethane elastomer.} \textbf{(a)} The experimentally measured average resilience of 6 polyurethane elastomer samples (left) and its simulated resilience (middle) both show a significant decrease when the loading and unloading rates are unequal. The resilience for the experimental timescales used (left and middle panels) is a part of a larger resilience profile (right) normalized by the mean characteristic relaxation time $\bar{\tau}$. The experimental timescales within this larger profile is denoted by the gray box and lies near the edge of the backbone region (yellow diagonal). \textbf{(b)} The simulations were performed using a generalized Maxwell solid model. The single spring $E_\infty$ represents the long-time equilibrium behavior of the solid, while each pair of spring and dashpot in series ($E_i$, $\tau_i$) represent a characteristic relaxation modulus and timescale in the material. \textbf{(c)} An $E_\infty$ of \SI{28.2}{MPa} and the values for $N = 90$ $E_i$, $\tau_i$ pairs (orange dots) for the spring-damper elements are obtained from a 1D linear viscoelastic characterization of the polyurethane elastomer. A Pearson distribution (blue curve) was fitted to the discrete relaxation spectrum to create a continuous representation of the spectrum.}
	\end{figure*}

	We measured the mechanical response of American bullfrog plantaris tendons (used to drive rapid motion during jumping) and guinea fowl lateral gastrocnemius tendons (used in walking and running). We stretched the tendons by increasing their length by 5\% and returning them back to their original length at controlled loading and unloading rates (Fig.~\ref{fig1}a-b). For equal loading and unloading rates, the tendons returned $\approx$ 80\% of the mechanical energy applied during stretching with a slight increase in resilience as the symmetric loading/unloading time increased (towards the upper-right in both Fig.~\ref{fig1}c-d). For asymmetric rates, we either applied a slow loading followed by a fast unloading or vice-versa. For the most extreme differences between the loading and unloading rates we measured, the resilience decreased of both types of tendons significantly to $\approx$ 60\% (Fig.~\ref{fig1}c-d). Given a fixed loading rate, we found that the maximum resilience generally occurs when the unloading rate matches the loading rate.
		
	To understand the generality and the underlying physical principles of this phenomenon, we performed similar measurements on synthetic elastomers. The synthetic elastomers enable us to measure their viscoelastic properties in the linear regime using time-temperature superposition that includes measurements on samples at sub-freezing temperatures, conditions which are inaccessible to the tendon samples without causing changes to the mechanical response of the tissue.
	
	Synthetic elastomer materials exhibit the same response as tendon to rate-asymmetry: they are most mechanically efficient when the loading and unloading rates are equal. The resilience of the polyurethane elastomer at a 1\% maximum strain (left panel of Fig.~\ref{fig2}a) was found to be relatively high with symmetric rates (\textgreater90\%), but had a $\approx$ 20\% drop in resilience over the two orders of magnitude in asymmetry measured. At higher maximum strains where nonlinear effects are more pronounced, the polyurethane elastomer showed a similar significant decrease in resilience with more asymmetry and was slightly less resilient overall (Supplemental Fig.~\ref{figS3}). A more dissipative elastomer, neoprene, was also consistent with this result (Supplemental Fig.~\ref{figS4}).
	
	To understand why materials are less energy efficient when subjected to asymmetric rates, we used the 1D generalized Maxwell model, a discrete linear viscoelastic representation of solids (Fig.~\ref{fig2}b). The model consists of an elastic element that accounts for the long-term equilibrium mechanical response of the material ($E_\infty$) in parallel with elastic and viscous (spring and dashpot) elements that each confer a stiffness and characteristic relaxation time ($E_i$ and $\tau_i$ for $i=1 \dots N$). We measured the linear viscoelastic response of the polyurethane elastomer using time-temperature superposition to obtain values for $E_\infty$ (\SI{28.2}{MPa}) and the discrete relaxation spectrum of $E_i$ versus $\tau_i$ (orange dots in Fig.~\ref{fig2}c) that we use to simulate the material response to symmetric and asymmetric loading and unloading rates. From this 1D linear viscoelastic model, the simulated resilience values are independent of the maximum strain (Supplemental Information: Independence from $\epsilon_\mathrm{max}$) and the modulus dependence is solely based on $E_0/E_\infty$, the ratio of the instantaneous modulus ($E_0 = E_\infty + \sum E_i$) to the equilibrium modulus ($E_\infty$), with a higher $E_0/E_\infty$ value corresponding to a more dissipative viscoelastic material (Supplemental Information: Dependence on $E_0/E_\infty$).

	\begin{figure*}[hbt]
		\centering
		\includegraphics[width=\linewidth]{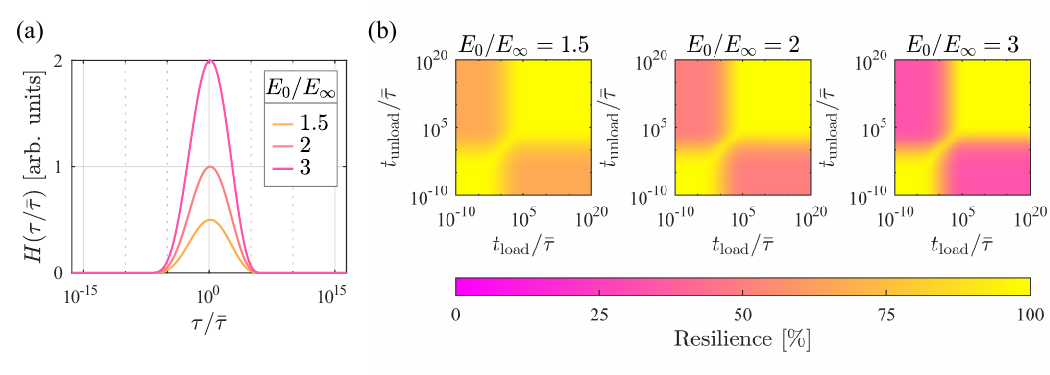}
		\caption{\label{fig3}\textbf{A more dissipative viscoelastic material has a lower resilience at asymmetric rates.} \textbf{(a)} Using the generalized Maxwell model, $E_0/E_\infty$ was varied from the experimental value of $E_0/E_\infty = 126$. \textbf{(b)} A larger value of $E_0/E_\infty$ (left to right on the panels), constituting a larger stiffness contribution from the spring-dashpot elements $\sum E_i$ compared to the long-term equilibrium spring $E_\infty$, lowers the resilience in the regions significantly impacted by asymmetric stretching (orange areas in left panel to pink areas in right panel) with the resilience converging to $E_\infty/E_0 \times 100\%$ as the rate-asymmetry is increased. The resilience at the extremes ($t_\mathrm{load},t_\mathrm{unload} << \bar{\tau}$ and $t_\mathrm{load},t_\mathrm{unload} >> \bar{\tau}$) depicted by the yellow squares is not significantly affected and remains at nearly 100\% resilience.}
	\end{figure*}
 
	To validate our modeling approach, we compared the simulated resilience to the measured resilience of the polyurethane elastomer under the same stretching conditions for the range of loading and unloading times tested. The simulated resilience shows an excellent agreement with the experimentally measured data (left and middle panels of Fig.~\ref{fig2}a) with a less than 2\% difference in resilience between the experiment and simulation (Supplemental Fig.~\ref{figS2}). This agreement between the experiment and simulation demonstrates that a simple 1D linear viscoelastic model captures the important mechanical properties necessary to reproduce the reduced efficiency at asymmetric rates.   
 
    Using the discrete relaxation spectrum and $E_\infty$ for the polyurethane elastomer, the resilience at timescales beyond the experimental loading/unloading times were simulated (right panel of Fig.~\ref{fig2}a) to capture the full resilience profile. We find regions of high resilience (yellow squares in the right panel of Fig.~\ref{fig2}a) for very short timescales (fast rates) and very long timescales (slow rates) relative to the mean characteristic relaxation time of the material $\bar{\tau}$. For the short timescale, none of the dissipative modes become active as long as $t_\mathrm{load} << \bar{\tau}$ and $t_\mathrm{unload} << \bar{\tau}$. For long timescales, the system is adiabatic and energy is approximately conserved in the material if $t_\mathrm{load} >> \bar{\tau}$ and $t_\mathrm{unload} >> \bar{\tau}$. In addition, we find a backbone region (the edge of which the experimental window is located) joining these extremes where the loading and unloading rates must closely match to have a high resilience (yellow diagonal in the right panel of Fig.~\ref{fig2}a). In the regions not covered by the extremes, a higher degree of asymmetry results in a lower resilience (pink areas in the right panel of Fig.~\ref{fig2}a).

    These regions of higher resilience do not necessarily correspond to a higher overall energy output. For short unloading times, the highest energy output occurs when the loading time is also short (Supplemental Fig.~\ref{figS5}), and the overall energy output of fast unloading decreases significantly when the loading time is increased (Supplemental Fig.~\ref{figS5}b). The elastic energy output can also be significantly modulated by changing the applied strain, or by changing the overall volume of elastic material. Because these factors are highly variable across different elastic systems, we focus on resilience rather than energy output in this work. 
 
	With the modeling approach validated, we further use a 1D linear viscoelastic model to explore how resilience depends on material properties over a broad range of loading and unloading rates. To reduce the number of parameters in our model, we represent the discrete relaxation spectrum as a continuous curve and fit a Pearson distribution on a $\log\tau$ scale (blue curve of Fig.~\ref{fig2}c). This fit depends on only four parameters: a mean of $\log\bar{\tau} = -8.56$, standard deviation of $\log\sigma = 2.05$, skewness of $\gamma_1 = -0.0847$, and kurtosis of $\beta_2 = 2.55$ (with $\gamma_1=0$ and $\beta_2 = 3$ for a Normal distribution). We use this continuous relaxation spectrum (blue curve of Fig.~\ref{fig2}c) with $N = 10^3$ discrete points ($E_i$, $\tau_i$) chosen equally spaced on the $\log\tau$ scale and the corresponding $E_0/E_\infty$ value of $126$ to match the polyurethane elastomer's linear viscoelastic properties as a starting point for the simulations. We find two primary properties of the material that shape the resilience response to asymmetric rates: (i) The ratio between the instantaneous modulus $E_0$ and the equilibrium modulus $E_\infty$ and (ii) the breadth of the material's relaxation spectrum characterized by the standard deviation $\log\sigma$.
 
	To explore how changes in $E_0$ relative to $E_\infty$ change the resilience, we compared the resilience profile of different $E_0/E_\infty$ values represented by a change in the height of the continuous relaxation spectrum with a constant $E_\infty$ (Fig.~\ref{fig3}a). At the short and long timescale extremes characteristic of high resilience (yellow squares in the panels of Fig.~\ref{fig3}b), we find that $E_0/E_\infty$ does not significantly affect the resilience. In contrast, the resilience is significantly impacted by $E_0/E_\infty$ in the regions of low resilience, asymmetric stretching (orange areas in the left panel to pink areas in the right panel of Fig.~\ref{fig3}b) with a lower resilience for larger values of $E_0/E_\infty$. The resilience in these regions converges to $E_\infty/E_0 \times 100\%$ as the loading and unloading rates become more asymmetric.

\begin{figure*}[htb]
		\centering
		\includegraphics[width=\linewidth]{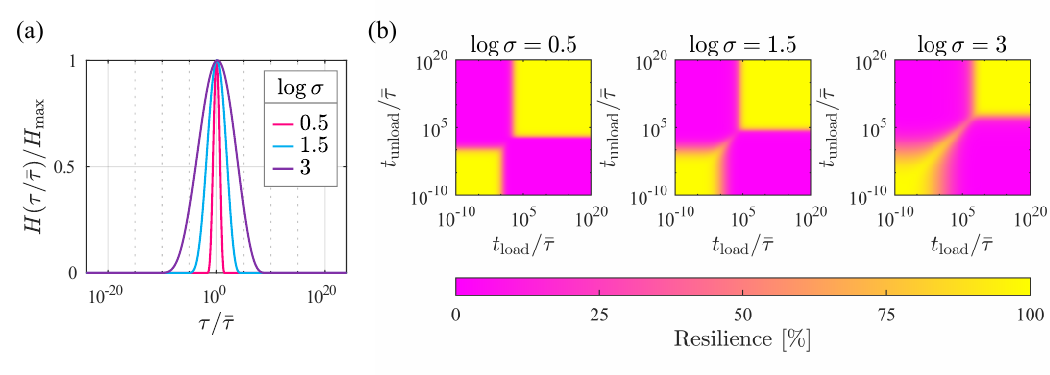}
		\caption{\label{fig4}\textbf{A broader viscoelastic relaxation spectrum increases the impact of rate-asymmetry on resilience.} \textbf{(a)} Using the generalized Maxwell model, we compare the effect of increasing the breadth of the relaxation spectrum by varying the standard deviation from the experimental value of $\log\sigma = 2.05$. \textbf{(b)} For a narrow relaxation spectrum (left panel), the material is resilient if both the loading and unloading rates are much slower or are both much faster than the mean characteristic relaxation time $\bar{\tau}$. For a broader relaxation spectrum (right panel), the loading and unloading rates need to closely match to have a high resilience.} 
	\end{figure*}

	Using the model, we also examined how changes to the distribution of the relaxation spectrum affect how a material responds to rate-asymmetry. By varying $\log\sigma$, we find that the breadth of the relaxation spectrum affects the resilience profile (Fig.~\ref{fig4}). For a material with a narrow distribution of relaxation times, the material is highly resilient if the loading and unloading times both occur on the very short or very long timescale extremes (left panel of Fig.~\ref{fig4}b). In these regions, the resilience is high even if the loading and unloading rates are quite different (yellow squares in the left panel of Fig.~\ref{fig4}b). However, for a broader relaxation spectrum, the resilience is high when the loading and unloading rates match, regardless of the experimental timescale (yellow diagonal in the right panel of Fig.~\ref{fig4}b). Thus, the breadth of the viscoelastic relaxation spectrum is important in shaping how the mechanical efficiency of the material responds to asymmetric rates. We also examined higher-order moments of the relaxation spectrum and found no significant changes with varying skewness (Supplemental Fig.~\ref{figS6}) while a higher kurtosis broadens the resilience transition between the long timescale extreme and asymmetric regions (Supplemental Fig.~\ref{figS7}).

    Finally, we verified the generality of the results to asymmetric loading and unloading conditions with non-constant rates. In the in-vivo tendon measurements (Fig.~\ref{fig0}), the strain-rate during each phase of movement is not constant. To explore this effect, we calculated the resilience for piecewise smooth sinusoidal strain loading and unloading profiles. The calculated resilience yields the same main result: over a broad range of rates viscoelastic materials recover elastic energy most efficiently when the loading and unloading rates match (Supplemental Fig.~\ref{figS8}).

	\section*{Discussion}

    Using a combination of experimental mechanical properties measurements and continuum material modeling, we find that viscoelastic materials are most mechanically efficient when loaded and unloaded at the same rate. The breadth of a material's viscoelastic relaxation spectrum is a key factor that determines the extent to which it will exhibit a drop in efficiency in response to asymmetric loading and unloading rates. Given the generality of our approach, we expect these results to apply to a variety of materials used in elastically-driven movements in biology such as resilin-rich cuticle in locusts~\cite{burrowsLocustsUseComposite2012} or chitinous exoskeleton used by mantis shrimp~\cite{patekComparativeSpringMechanics2013} as well as the actuation of soft elastomers in recent robotics applications~\cite{chi2022snapping,kim2023laser}. 
       
    For biological materials with a broad distribution of relaxation times, various levels of the hierarchical material structure could respond differently to asymmetry in the loading and unloading rates. The materials that comprise biological springs are often composite materials that yield a complex mechanical response to deformation~\cite{burrows2008resilin,patek2011bouncy}. Biological materials have a multi-scale hierarchical structure that causes multiple important length-scales and timescales to emerge in response to mechanical deformation~\cite{bao2003cell,buehler2007nano,fung2013biomechanics,tang2021dynamic}. This hierarchical structure results in a broad viscoelastic relaxation spectrum caused by the different dynamical properties at the individual protein level~\cite{bao2003cell} and at the whole tissue level~\cite{fung2013biomechanics,tang2021dynamic}.

    In our measurements here, we are operating at the tissue-level and see pronounced differences with a $\sim$ 20\% drop in resilience at asymmetric rates. At the next level down from the whole tissue, tendons are comprised of tendon fascicles. Rosario \& Roberts examined the mechanical properties of rat tail tendon fascicles and observed no dependence on rate-asymmetry~\cite{rosario2020loading}. This suggests that perhaps the important broadening of the relaxation spectrum of tendon for its rate-asymmetric behavior occurs at the tissue-level. However, there are some important limitations to this comparison. The fascicle measurements were performed at close to symmetric rates (only a factor of 0.5-4x difference in loading/unloading rate), the lower forces of fascicles yielded noisier data, and there are functional differences between rat tail tendon and the energy storing tendons we examined in this work. Future work comparing the relaxation spectrum across multiple levels of biological material structure would provide insight into the emergence of rate-asymmetry. Structural differences found in energy storing versus positional tendons~\cite{shearer2017relative} may also reveal differences in the rate-asymmetric response of tendons.

    Our results here also have potential impact in comparative biomechanics. Previous work has highlighted how strain rate can affect functional traits of biomechanical systems~\cite{anderson2022different}. Translating our results here into the language of ref~\cite{anderson2022different}: the viscoelastic relaxation spectrum of the tendon (functional trait) determines the tendon's resilience (attribute) based on the loading/unloading rate of the material (external factor). Future work could probe deeper into this structure-property relationship of tendon. Previous work has demonstrated there are morphological changes that correlate to changes in external factors or overall function~\cite{aziziBiaxialStrainVariable2009,abdalaFrogTendonStructure2018,lieberFrogSemitendinosisTendon1991}. Our work here provides a framework for future investigation into these morphological changes. Measuring specific properties of the viscoelastic functional trait of biological materials (specifically $E_0/E_\infty$ and the breadth of the spectrum) determines how efficiently they will respond to a variety of external loading factors. Therefore, connecting changes in morphology to these specific properties could inform the analysis of form-function relationships in biomechanical systems. 

    In animals that use latch-mediated spring actuation, muscles cannot load elastic energy quickly enough to match the unloading rates, otherwise the elastic structure would not be required to have a high kinematic output. However, there is some evidence that these timescales are not on the extremes of asymmetry (e.g. smasher mantis shrimp still load at \SI{300}{ms} even though they could take longer because of their slow-moving prey)~\cite{patekPowerMantisShrimp2019}. Why don't these animals use slower muscle contractions to load their elastic mechanisms? From our work here, we see there would be no energetic benefit to loading more slowly. A slower loading would increase the rate-asymmetry and thus decrease the resilience of the elastic materials. For a short unloading time like those characteristic of ultra-fast movements, a shorter loading time actually yields an overall higher energy output (Supplemental Fig.~\ref{figS5}) due to stress-relaxation that occurs during a slow loading. 

    Biological springs are also used for damping and deceleration (power attenuation) in landing and declined running movements~\cite{robertsFlexibleMechanismsDiverse2011}. The performance of elastic structures in biomechanics will often need to balance multiple roles, such as balancing energy conservation and damping in passive elastic movements found in cockroach legs~\cite{dudek2006passive}. Our work here suggests that viscoelastic materials will have a similar energy efficiency when used during power attenuating movements (fast loading followed by a slow unloading) compared to latch-mediated spring actuation movements (slow loading followed by fast unloading). Future work measuring the asymmetric rate response of biological materials that balance different functional roles might uncover interesting distinctions between them. 

    Our work suggests that symmetric loading and unloading rates are the most mechanically efficient, but gaits that involve cyclic loading and unloading can be biased towards a shorter loading time. For example, asymmetric ground reaction forces have been observed in running birds and lizards, with the loading phase occurring more rapidly than unloading~\cite{daley2003muscle,birn2014don,clemente2018steady}. This asymmetric gait appears to be more pronounced among animals that have a body CoM positioned forward of the hip~\cite{clemente2018steady}. However, humans also exhibit asymmetry in the elastic bounce of the leg and body, observed in ground reaction forces and body CoM motion, but the asymmetry decreases with increasing speed~\cite{nilsson1989ground,cavagna2010symmetry}. Models that include intrinsic damping in the leg with a single characteristic relaxation time predict these asymmetric gaits are energetically optimal~\cite{birn2014don}. But our results here suggest that a broad spectrum of relaxation times could be important for the emergence of symmetric rate efficiency, and biomechanical models that incorporate a more realistic viscoelastic model could uncover shifts in energetically optimal gaits. 
   
    There are some important limitations in the scope of this work. Although here we focused on the efficiency of energy released from elastic structures, there are other behavioral and physiological factors that contribute to the relative loading and unloading rates of elastic biomechanical systems. The dynamics of the entire biomechanical system contribute to the overall performance, with muscle properties, linkage mechanics, and load mass all being important for the kinematics of fast elastic movements~\cite{iltonPrinciplesCascadingPower2018}. Energy can be lost during muscle contractions and to the environment, and there are multiple ways that animals minimize the energetic cost of locomotion. For example, changing gait and muscle activation patterns during running can help minimize losses during ground contact~\cite{rebula2015cost,polet2019inelastic} and increase efficiency during incline running~\cite{daley2003muscle}.

    Other limitations in the work include simplifications made in the model. We used a 1D linear viscoelastic model as a minimal model to capture the essential features of the asymmetric rate effect. But there are multi-axial stress states and non-linear viscoelasticity in tendons~\cite{cheng2009mechanical}. We also considered tendons and synthetic viscoelastic materials of a relatively similar size, but the size-scale of the elastic material can also be an important consideration when assessing its kinematic performance~\cite{iltonEffectSizescaleKinematics2019}. At small-size scales viscous drag forces can significantly impede locomotion~\cite{bennet1975scale} and can be a significant source of dissipation for fast elastic movements~\cite{challita2021slingshot}. These additional effects could be added to extend the model and further generalize the results of this work.

    \section*{Conclusion}

     In this study, we examined the mechanical energy efficiency of biological springs and synthetic materials under symmetric and asymmetric loading/unloading rates. Our results reveal that these materials are most mechanically efficient when subjected to symmetric loading and unloading rates. This finding holds true for both tendons and synthetic elastomers. By utilizing a 1D linear viscoelastic model to analyze the data, we gained insight into the underlying principles governing the resilience of these materials. We found that the ratio between the instantaneous modulus and the equilibrium modulus as well as the breadth of the viscoelastic relaxation spectrum influence the impact of rate-asymmetry on mechanical efficiency. These insights can be applied to various fields, including soft robotics and bioengineering, which can benefit from efficient elastic mechanisms. By understanding how materials respond to different loading and unloading rates, the design of these systems can be optimized for maximum performance.

\section*{Materials and Methods}

\subsubsection*{Tendons}
	
	Bullfrog plantaris tendons were extracted from previously frozen cadavers of the American bullfrog (\textit{Lithobates catesbeianus}). The plantaris muscle-tendon unit was isolated from the tibiafibula. Muscle tissue was then carefully removed from the aponeurosis and tendon. The elastic tissue was kept moist with physiological saline before being frozen again for use in experiments. Previous work has shown that given the low cellular density in tendon, freezing has no effect on the general mechanical properties of tendons~\cite{LeeElliott2017}. All procedures were approved by the University of California, Irvine Animal Care and Use Committee (protocol AUP 20-129).
 
 	Guinea fowl common gastrocnemius tendons were extracted from previously frozen cadavers of \textit{Numida meleagris}. The tendon was carefully removed from the insertion point at the tarsometatarsus and the muscle-tendon unit was carefully dissected out and removed from its origin on the femur. Muscle tissue was carefully removed, and the tendon tissue was kept moist with physiological saline before being frozen again to store for later experiments. All procedures were approved by the University of California, Irvine Animal Care and Use Committee (protocol AUP-23-031).

	The tendons were removed from the freezer and thawed in a saline solution before being epoxied (J-B Weld WaterWeld Epoxy Putty) between a pair of 3D printed rectangular plates on both ends or clamped between a pair of laser cut acrylic plates with peak-to-peak serrations to prevent slippage~\cite{SHI2012516, Cheung2006ASJ, Jiang2020ClampingSB}. The resting length of the tendons were \SIrange{15}{22}{mm} between the printed plates, with a thickness of \SIrange{0.5}{1.5}{mm}. The width of the tendons at the edge of the printed plates was \SIrange{10}{20}{mm}.
	
	\subsubsection*{Synthetic elastomers}
 
	Prefabricated synthetic elastomer sheets were obtained from McMaster-Carr (polyurethane: catalog \#2178T32, thickness $1/32$ in, 90A durometer; neoprene: catalog \#1370N32, thickness $1/32$ in, 60A durometer). These sheets were sectioned into rectangular strips \SI{6}{mm} by \SI{40}{mm}.
	
	\subsection*{Custom-waveform DMA}
 
	Using the arbitrary waveform mode of the TA Instruments RSA-G2 dynamic mechanical analyzer (DMA), symmetric and asymmetric triangular waves were constructed by superimposing positively sloped and negatively sloped strain versus time lines for the loading and unloading phase respectively. Combinations of five loading times (\SI{0.1}{s}, \SI{0.316}{s}, \SI{1}{s}, \SI{3.16}{s}, and \SI{10}{s}) and five unloading times (\SI{0.1}{s}, \SI{0.316}{s}, \SI{1}{s}, \SI{3.16}{s}, and \SI{10}{s}) were used to create 25 different loading and unloading combinations. The results in Supplemental Table S1-S2 highlight the loading and unloading time combinations used. The times were chosen to span a difference of two orders of magnitude between the shortest and longest time. 

    The samples were secured to the tensile clamps of the DMA and a randomized ordering of the 25 different loading and unloading combinations were conducted at $\sim$ \SI{23}{\celsius} for the selected maximum strain. Three cycles were performed consecutively at a single loading/unloading combination. Between each loading/unloading combination, the samples were allowed to relax \SI{120}{s} at zero strain to ensure the previous loading history of the material did not affect the results. For the synthetic elastomers, the loading gap with the clamped sample was increased incrementally so that the sample experienced no initial compressive stress. For the tendon samples, $\sim$ \SI{0.05}{N} of preload force was imposed to reach a sensitive force range that resulted in a final loading gap of $\sim$ \SI{18}{mm}. A preconditioning protocol of 12 cycles at 3\% strain was applied to the tendon samples before testing, consistent with the recommendations of ref. ~\cite{ebrahimi2019effect}. To keep the tendon samples hydrated, saline solution was applied using a cotton swab during each relaxation period. Because we measured tensile force for thin samples near their equilibrium length, we did not measure any significant negative force due to buckling of the material in response to compression.

    The energy input and output for a cycle were computed by numerically integrating the positive area under the stress versus strain curve for the loading and unloading phase respectively. Resilience was calculated by dividing the energy output by the energy input at the third loading and unloading cycle to mitigate transient material response and the Mullins effect~\cite{cheng2009mechanical}. 
	
    \subsection*{Modeling the Linear Viscoelastic Resilience of Elastomers}
 
    To characterize the linear viscoelastic properties of the synthetic polyurethane elastomer, a frequency sweep from \SI{1}{Hz} to \SI{15}{Hz} at an oscillation amplitude of \SI{1.0}{N} was applied to a sample using a TA Instruments Q800 dynamic mechanical analyzer. The frequency sweep was first conducted at \SI{-40}{\celsius} and then in increasing temperature increments of \SI{5}{\celsius} up to \SI{55}{\celsius}. In the TA Instruments TRIOS software, the TTS functionality was used to shift the storage modulus versus frequency data from all of the temperatures in the experiment to \SI{23}{\celsius}, the approximate temperature in which the symmetric and asymmetric stretching tests were performed (Supplemental Fig.~\ref{figS1}). With a focus on the glassy-rubbery transition, the values of storage modulus at the lowest and highest frequency were extended beyond the experimental shifted frequencies to elongate the glassy and rubbery plateau regions. A spline fit interpolation was used to smooth the storage modulus (yellow curve of Supplemental Fig.~\ref{figS1}c). The relaxation modulus, $E_\mathrm{PU}(t)$, for the polyurethane elastomer was obtained from the sine transform of the storage modulus (Supplemental Equation S1).

    To model the resilience of elastomers in the linear viscoelastic regime, the generalized Maxwell model consisting of a spring (stiffness $E_\infty$) in parallel with $N$ elements of springs (stiffness $E_i$) and dashpots (characteristic relaxation time $\tau_i$) in series was used, where $i=1 \dots N$ (Fig.~\ref{fig2}b). In the generalized Maxwell model, the relaxation modulus is represented by the equation~\cite{tschoegl2012phenomenological}
    \begin{equation}  \label{eq:Gen_Maxwell}
        E(t) = E_\infty + \sum_{i=1}^N E_i e^{-t/\tau_i}.
    \end{equation}
    
    For the polyurethane elastomer, $E_\infty$ was obtained from the equilibrium value of $E_\mathrm{PU}(t)$. To determine the values of $E_i$ and $\tau_i$, a sum of $N$ exponential decays ($E_i e^{-t/\tau_i}$) with equally spaced values for $\tau_i$ on the $\log\tau$ scale across the range of times used for the relaxation modulus was fitted to $E_\mathrm{PU}(t) - E_\infty$ using the Levenberg-Marquardt algorithm (Supplemental Fig.~\ref{figS1}d). From the sum of squared residuals of the fit across a range of $N$ values, $N = 90$ spring-dashpot elements were chosen from the convergence of the sum of squared residuals for a parsimonious fit (Supplemental Fig.~\ref{figS1}e). From the values for $E_i$ versus $\tau_i$ (representing a discrete relaxation spectrum) and $E_\infty$, the stress response from different loading and unloading combinations were simulated, and the resilience at the third cycle was computed for the polyurethane elastomer (Supplemental Equation S13).

    \section*{Acknowledgements}

    The authors thank Mikayla Mann for helpful discussions. M.I. acknowledges funding support from the NSF for this work under grant no. 2019371. L.T., P.N., I.W., and T.L. acknowledge funding support from the Harvey Mudd College Physics Summer Research Fund.   

    \section*{Author Contributions}
    \textbf{Lucien Tsai:} Conceptualization, Methodology, Software, Investigation, Writing - Original Draft, Writing - Review \& Editing, Visualization. \textbf{Paco Navarro:} Validation, Investigation, Writing - Review \& Editing. \textbf{Siqi Wu:} Investigation, Methodology, Writing - Review \& Editing. \textbf{Taylor Levinson:} Investigation, Methodology, Writing - Review \& Editing. \textbf{Elizabeth Mendoza:} Investigation, Writing - Review \& Editing. \textbf{M. Janneke Schwaner:} Investigation, Writing - Review \& Editing.
    \textbf{Monica A. Daley:} Conceptualization, Resources, Writing - Review \& Editing
    \textbf{Emanuel Azizi:} Conceptualization, Resources, Writing - Original Draft, Writing - Review \& Editing, 
    Visualization
    \textbf{Mark Ilton:} Conceptualization, Resources, Writing - Original Draft, Writing - Review \& Editing, Supervision. 

	\bibliography{posmlab}

\newpage

\onecolumngrid
\renewcommand{\thefigure}{S\arabic{figure}}
\setcounter{figure}{0}

\renewcommand{\theequation}{S\arabic{equation}}
\setcounter{equation}{0}

\appendix

\section*{Supplemental Information}

\section*{Relaxation Modulus Transformation}
The relaxation modulus $E(t)$ is transformed from the storage modulus $E'(w)$ in frequency-space by the equation~\cite{christensen2012theory}
\begin{equation} \label{eq:Modulus}
    E(t) = \frac{2}{\pi}\int_0^\infty \frac{E'(w)}{w}\sin(wt) dw.
\end{equation}

\section*{Generalized Maxwell Model \& Resilience Computation: Linear Loading \& Unloading}

The stress response of a single spring-damper series, or Maxwell element, in the generalized Maxwell model is first considered. Combined with the $E_\infty$ spring, the stress response of the entire generalized Maxwell model is derived, from which the resilience is numerically determined.

\subsection*{Maxwell Element Stress-Strain Relationship}

For the $i$th Maxwell element consisting of a spring ($E_i$) in series with a dashpot ($\tau_i$), the constitutive differential equation is given by the equation~\cite{tschoegl2012phenomenological}
\begin{equation} \label{eq:DE}
    \dot{\sigma} + \frac{\sigma}{\tau_i} = E_i\dot{\epsilon}.
\end{equation}

\subsubsection*{Loading Phase}

During the loading phase, the strain versus time function is given by
\begin{equation} \label{eq:loading_strain}
    \epsilon(t) = \frac{\epsilon_\mathrm{max}}{t_\mathrm{load}}t
\end{equation}
for $0 \leq t \leq t_\mathrm{load}$, where $\epsilon_\mathrm{max}$ is the maximum strain and $t_\mathrm{load}$ is the loading time. \\

By solving the differential equation in Equation (\ref{eq:DE}) for $\sigma(t)$ with the imposed $\epsilon(t)$ function in Equation (\ref{eq:loading_strain}), the stress response during the loading phase at the $j$th cycle is
\begin{equation} \label{eq:loading_stress}
    \sigma_{i,j,\mathrm{load}}(t) = \frac{E_i \tau_i \epsilon_\mathrm{max}}{t_\mathrm{load}} + \left[\sigma_{i,j-1,\mathrm{unload}}(t_\mathrm{unload}) - \frac{E_i \tau_i \epsilon_\mathrm{max}}{t_\mathrm{load}}\right]e^{-t/\tau_i},
\end{equation}
where $\sigma_{i,j-1,\mathrm{unload}}(t_\mathrm{unload})$ is the stress at the start of the $j$th cycle for this single Maxwell element and $\sigma_{i,0,\mathrm{unload}}(t_\mathrm{unload}) = 0$ for the condition of no initial stress. \\

Since $\epsilon(t) = \frac{\epsilon_\mathrm{max}}{t_\mathrm{load}}t \Rightarrow t = \frac{t_\mathrm{load} \epsilon}{\epsilon_\mathrm{max}}$, the stress response can be reparameterized into
\begin{equation} \label{eq:loading_stress_reparam}
    \sigma_{i,j,\mathrm{load}}(\epsilon) = \frac{E_i \tau_i \epsilon_\mathrm{max}}{t_\mathrm{load}} + \left[\sigma_{i,j-1,\mathrm{unload}}(0) - \frac{E_i \tau_i \epsilon_\mathrm{max}}{t_\mathrm{load}}\right]e^{-\frac{t_\mathrm{load}}{\tau_i}\frac{\epsilon}{\epsilon_\mathrm{max}}}.
\end{equation}
     
\subsubsection*{Unloading Phase}
During the following unloading phase, the strain versus time function is given by
\begin{equation} \label{eq:unloading_strain}
    \epsilon(t) = \epsilon_\mathrm{max} - \frac{\epsilon_\mathrm{max}}{t_\mathrm{unload}}t
\end{equation}
for $0 \leq t \leq t_\mathrm{unload}$, where $t_\mathrm{unload}$ is the unloading time. \\

By solving the differential equation in Equation (\ref{eq:DE}) for $\sigma(t)$ with the imposed $\epsilon(t)$ function in Equation (\ref{eq:unloading_strain}), the stress response during the unloading phase at the $j$th cycle is
\begin{equation} \label{eq:unloading_stress}
    \sigma_{i,j,\mathrm{unload}}(t) = -\frac{E_i \tau_i \epsilon_\mathrm{max}}{t_\mathrm{unload}} + \left[\sigma_{i,j,\mathrm{load}}(t_\mathrm{load}) + \frac{E_i \tau_i \epsilon_\mathrm{max}}{t_\mathrm{unload}}\right]e^{-t/\tau_i},
\end{equation}
where $\sigma_{i,j,\mathrm{load}}(t_\mathrm{load})$ is the stress at the end of the preceding loading phase at the $j$th cycle for this single Maxwell element. \\

Since $\epsilon(t) = \epsilon_\mathrm{max} - \frac{\epsilon_\mathrm{max}}{t_\mathrm{unload}}t, \Rightarrow t = t_\mathrm{unload}\left(\frac{\epsilon}{1 - \epsilon_\mathbf{max}}\right)$, the stress response can be reparameterized into
\begin{equation} \label{eq:unloading_stress_reparam}
    \sigma_{i,j,\mathrm{unload}}(\epsilon) = -\frac{E_i \tau_i \epsilon_\mathrm{max}}{t_\mathrm{unload}} + \left[\sigma_{i,j,\mathrm{load}}(\epsilon_\mathrm{max}) + \frac{E_i \tau_i \epsilon_\mathrm{max}}{t_\mathrm{unload}}\right]e^{-\frac{t_\mathrm{unload}}{\tau_i}\left(1 - \frac{\epsilon}{\epsilon_\mathrm{max}}\right)}.
\end{equation}

\subsection*{Generalized Maxwell Model Stress-Strain Relationship}

The generalized Maxwell model consists of a spring of stiffness $E_\infty$ in parallel with $N$ Maxwell elements. Thus, the total stress response is the sum of the stress responses from the spring with stiffness $E_\infty$ and each Maxwell element. Since the stress response from the spring with stiffness $E_\infty$ is $\sigma_\infty(\epsilon) = E_\infty \epsilon$, the stress-strain relationship during the loading phase at the $j$th cycle for the generalized Maxwell model consisting of $N$ Maxwell elements is given by
\begin{equation} \label{eq:Gen_loading_sress}
\begin{split}
    \sigma_{j,\mathrm{load}}(\epsilon) &= \sigma_\infty(\epsilon) + \sum_{i=1}^N \sigma_{i,j,\mathrm{load}}(\epsilon) \\
    &= E_\infty \epsilon + \sum_{i=1}^N \left(\frac{E_i \tau_i \epsilon_\mathrm{max}}{t_\mathrm{load}} + \left[\sigma_{i,j-1,\mathrm{unload}}(0) - \frac{E_i \tau_i \epsilon_\mathrm{max}}{t_\mathrm{load}}\right]e^{-\frac{t_\mathrm{load}}{\tau_i}\frac{\epsilon}{\epsilon_\mathrm{max}}}\right).
\end{split}
\end{equation}

Similarly, the stress-strain relationship during the unloading phase at the $j$th cycle is given by
\begin{equation} \label{eq:Gen_unloading_stress}
\begin{split}
    \sigma_{j,\mathrm{unload}}(\epsilon) &= \sigma_\infty(\epsilon) + \sum_{i=1}^N \sigma_{i,j,\mathrm{unload}}(\epsilon) \\
    &= E_\infty \epsilon + \sum_{i=1}^N \left(-\frac{E_i \tau_i \epsilon_\mathrm{max}}{t_\mathrm{unload}} + \left[\sigma_{i,j,\mathrm{load}}(\epsilon_\mathrm{max}) + \frac{E_i \tau_i \epsilon_\mathrm{max}}{t_\mathrm{unload}}\right]e^{-\frac{t_\mathrm{unload}}{\tau_i}\left(1 - \frac{\epsilon}{\epsilon_\mathrm{max}}\right)}\right).
\end{split}
\end{equation}

\subsection*{Resilience Computation}

The energy stored during the $j$ cycle is given by
\begin{equation} \label{eq:Ein}
    E_\mathrm{in,j} = \int_0^{\epsilon_\mathrm{max}} \max(\sigma_{j,\mathrm{load}}(\epsilon),0) d\epsilon.
\end{equation}
The energy released during the $j$ cycle is given by
\begin{equation} \label{eq:Eout}
    E_\mathrm{out,j} = \int_0^{\epsilon_\mathrm{max}} \max(\sigma_{j,\mathrm{unload}}(\epsilon),0) d\epsilon.
\end{equation}
The resilience, $\Gamma$, at the $j$th cycle is then
\begin{equation} \label{eq:resilience}
    \Gamma = \frac{E_\mathrm{out,j}}{E_\mathrm{in,j}} \times 100\%.
\end{equation}

\subsubsection*{Dependence on $E_0/E_\infty$}

For the stress response of a single Maxwell element (Equations~\eqref{eq:Gen_loading_sress} and \eqref{eq:Gen_unloading_stress}) from loading and unloading at any cycle, $E_i$ only appears as a common coefficient for each term. In addition, $E_\infty$ only appears as the coefficient of the single term in the stress response of the $E_\infty$ spring. Thus, if each modulus term is multiplied by a positive real constant $a$, the new stress response from the entire generalized Maxwell model is scaled by a factor of $a$ from the original stress response when $a = 1$.

Additionally, the new resilience (Equation~\eqref{eq:resilience}) is the same as the original resilience since the constant term $a$ is removed from the division of $E_\mathrm{out,j}$ over $E_\mathrm{in,j}$.

Since $\sum aE_i$ and $aE_\infty$ yields the same resilience as $\sum E_i$ and $E_\infty$, the modulus dependence of resilience is solely based on the ratio of $\left(\sum E_i\right)/E_\infty$ or $E_0/E_\infty$ with $E_0 = E_\infty + \sum E_i$.

\subsubsection*{Independence from $\epsilon_\mathrm{max}$}

The energy stored and released in Equations~\eqref{eq:Ein} and \eqref{eq:Eout} can be rewritten as
\begin{equation} \label{eq:Ein_mod}
    E_\mathrm{in,j} = \int_{\alpha\epsilon_\mathrm{max}} ^{\epsilon_\mathrm{max}} \sigma_{j,\mathrm{load}}(\epsilon) d\epsilon
\end{equation}
and
\begin{equation} \label{eq:Eout_mod}
    E_\mathrm{out,j} = \int_{\beta\epsilon_\mathrm{max}}^{\epsilon_\mathrm{max}} \sigma_{j,\mathrm{unload}}(\epsilon) d\epsilon,
\end{equation}
respectively, where $\alpha\epsilon_\mathrm{max}$ is the strain at which the loading curve crosses $\sigma = 0$ ($\alpha = 0$ for the first loading cycle) and $\beta\epsilon_\mathrm{max}$ is the strain at which the unloading curve crosses $\sigma = 0$.

By performing the integration in Equations~\eqref{eq:Ein_mod} and \eqref{eq:Eout_mod}, the sole dependence on $\epsilon_\mathrm{max}$ for $E_\mathrm{in,j}$ and $E_\mathrm{out,j}$ is the coefficient $\epsilon_\mathrm{max}^2$ in every term. Since, $\epsilon_\mathrm{max}^2$ is removed from the division of $E_\mathrm{out,j}$ over $E_\mathrm{in,j}$ to compute resilience (Equation~\eqref{eq:resilience}), resilience is independent of $\epsilon_\mathrm{max}$.

\section*{Generalized Maxwell Model \& Resilience Computation: Harmonic Loading \& Unloading}

The same method to compute the resilience for linear loading \& unloading is applied to harmonic loading \& unloading, where the loading and unloading curves consist of piecewise smooth sinusoidal functions with different loading \& unloading times. These harmonic functions yield a different form for the stress response compared to the linear (constant strain-rate) loading \& unloading. The stress response for the $i$th Maxwell element under harmonic loading \& unloading is derived in this section.

\subsubsection*{Loading Phase}

During the loading phase, the strain versus time function is given by
\begin{equation} \label{eq:Harmonic_loading_strain}
    \epsilon(t) = \frac{\epsilon_\mathrm{max}}{2} - \frac{\epsilon_\mathrm{max}}{2}\cos{\left(\frac{\pi t}{t_\mathrm{load}}\right)}
\end{equation}
for $0 \leq t \leq t_\mathrm{load}$, where $\epsilon_\mathrm{max}$ is the maximum strain and $t_\mathrm{load}$ is the loading time. \\

Taking the Laplace transform of Equation (\ref{eq:DE}) with the imposed $\epsilon(t)$ function in Equation (\ref{eq:Harmonic_loading_strain}), the stress response during the loading phase at the $j$th cycle is
\begin{equation} \label{eq:Harmonic_loading_stress_original}
    \Sigma_{i,j,\mathrm{load}}(s) = \frac{E_i \epsilon_\mathrm{max} \pi^2}{2 t_\mathrm{load}^2}\frac{1}{(s^2 + \pi^2/t_\mathrm{load}^2)(s + 1/\tau_i)} + \frac{\sigma_{i,j-1,\mathrm{unload}}(t_\mathrm{unload})}{s + 1/\tau_i},
\end{equation}
where the Laplace transform of $\sigma(t)$ is $\Sigma(s)$, $\sigma_{i,j-1,\mathrm{unload}}(t_\mathrm{unload})$ is the stress at the start of the $j$th cycle for this single Maxwell element, and $\sigma_{i,0,\mathrm{unload}}(t_\mathrm{unload}) = 0$ for the condition of no initial stress. \\

Using partial fraction decomposition,
\begin{equation} 
    \Sigma_{i,j,\mathrm{load}}(s) = \frac{E_i \epsilon_\mathrm{max}}{2}\frac{1}{1 + t_\mathrm{load}^2/\pi^2 \tau_i^2}\left(-\frac{s - 1/\tau_i}{s^2 + \pi^2/t_\mathrm{load}^2} + \frac{1}{s + 1/\tau_i}\right) + \frac{\sigma_{i,j-1,\mathrm{unload}}(t_\mathrm{unload})}{s + 1/\tau_i}.
\end{equation}

Performing the inverse Laplace transform,
\begin{equation} \label{eq:Harmonic_loading_stress}
    \sigma_{i,j,\mathrm{load}}(s) = \frac{E_i \epsilon_\mathrm{max}}{2}\frac{1}{1 + t_\mathrm{load}^2/\pi^2 \tau_i^2}\left(\frac{t_\mathrm{load}}{\pi \tau_i}\sin{\left(\frac{\pi t}{t_\mathrm{load}}\right)} - \cos{\left(\frac{\pi t}{t_\mathrm{load}}\right)} + e^{-t/\tau_i}\right) + \sigma_{i,j-1,\mathrm{unload}}(t_\mathrm{unload})e^{-t/\tau_i}.
\end{equation}

\subsubsection*{Unloading Phase}
During the following unloading phase, the strain versus time function is given by
\begin{equation} \label{eq:Harmonic_unloading_strain}
    \epsilon(t) = \frac{\epsilon_\mathrm{max}}{2} + \frac{\epsilon_\mathrm{max}}{2}\cos{\left(\frac{\pi t}{t_\mathrm{unload}}\right)}
\end{equation}
for $0 \leq t \leq t_\mathrm{unload}$, where $t_\mathrm{unload}$ is the unloading time. \\

Taking the Laplace transform of Equation (\ref{eq:DE}) with the imposed $\epsilon(t)$ function in Equation (\ref{eq:Harmonic_unloading_strain}), the stress response during the unloading phase at the $j$th cycle is
\begin{equation} 
    \Sigma_{i,j,\mathrm{unload}}(s) = -\frac{E_i \epsilon_\mathrm{max} \pi^2}{2 t_\mathrm{unload}^2}\frac{1}{(s^2 + \pi^2/t_\mathrm{unload}^2)(s + 1/\tau_i)} + \frac{\sigma_{i,j,\mathrm{load}}(t_\mathrm{load})}{s + 1/\tau_i},
\end{equation}
where the Laplace transform of $\sigma(t)$ is $\Sigma(s)$ and $\sigma_{i,j,\mathrm{load}}(t_\mathrm{load})$ is the stress at the end of the preceding loading phase at the $j$th cycle for this single Maxwell element.

Noting that $\Sigma_{i,j,\mathrm{unload}}(s)$ has the same form as Equation (\ref{eq:Harmonic_loading_stress_original}), the stress is given by
\begin{equation} 
    \sigma_{i,j,\mathrm{unload}}(s) = \frac{E_i \epsilon_\mathrm{max}}{2}\frac{1}{1 + t_\mathrm{unload}^2/\pi^2 \tau_i^2}\left(-\frac{t_\mathrm{unload}}{\pi \tau_i}\sin{\left(\frac{\pi t}{t_\mathrm{unload}}\right)} + \cos{\left(\frac{\pi t}{t_\mathrm{unload}}\right)} - e^{-t/\tau_i}\right) + \sigma_{i,j,\mathrm{load}}(t_\mathrm{load})e^{-t/\tau_i}.
\end{equation}

\clearpage

\begin{figure}
\centering
\includegraphics[width=\textwidth]{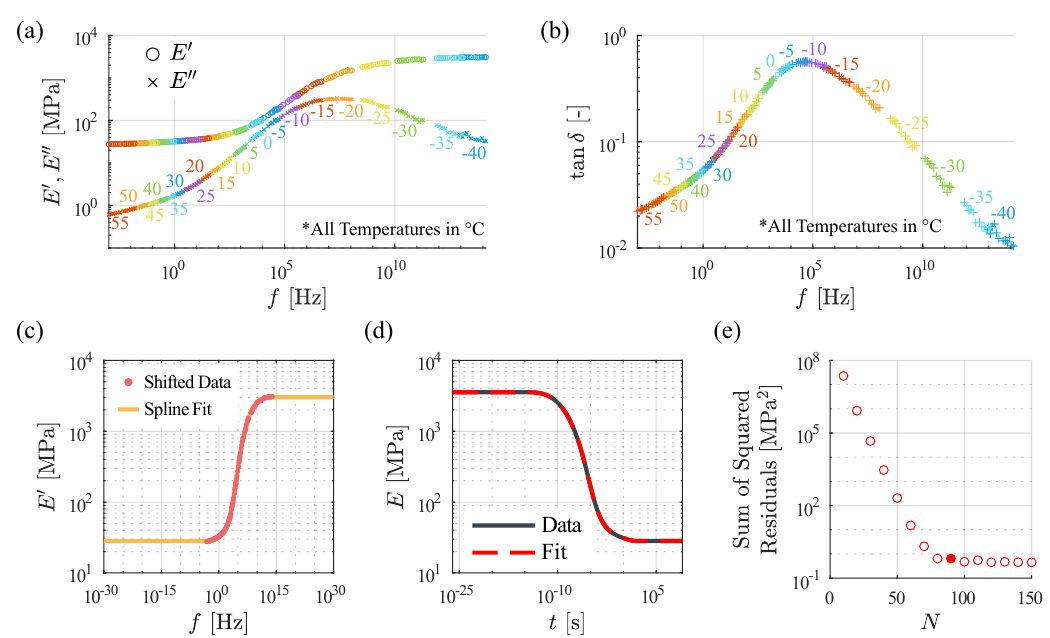}
\caption{\label{figS1}\textbf{The discrete relaxation spectrum is obtained from the storage modulus of the polyurethane elastomer.} \textbf{(a)} The storage ($E'$) and loss ($E''$) modulus for the polyurethane elastomer is obtained using time-temperature superposition (TTSP) and are shifted to \SI{23}{\celsius} in frequency. The numbers along the loss modulus curve represent the temperature at which the frequency sweep was conducted. \textbf{(b)} The corresponding $\tan\delta$ to panel (a) for the polyurethane elastomer. \textbf{(c)} The value of the shifted storage modulus $E'$ (orange dots) at the smallest and largest frequency is extended to form the glassy and rubbery plateau regions. A spline fit is performed to smooth the storage modulus (yellow curve). \textbf{(d)} The relaxation modulus (black curve) is obtained from the sine transform of the spline-fitted storage modulus using Equation (\ref{eq:Modulus}). Using the general Maxwell model, the relaxation modulus is fitted to Equation (\ref{eq:Gen_Maxwell}) with $N = 90$ spring-dashpot elements using the Levenberg-Marquardt algorithm to obtain a fit (red dashed curve) that matches well with the relaxation modulus data. From this fit, the values of $E_i$, $\tau_i$ pairs for the spring-dashpot elements are obtained to form the discrete relaxation spectrum (orange dots of Fig.~\ref{fig2}c) \textbf{(e)} The number of spring-dashpot elements, $N$, is varied and the corresponding sum of squared residuals is computed from the fit to the relaxation modulus data. $N = 90$ (solid red dot) spring-dashpot elements are chosen based on the convergence of the sum of squared residuals while larger values for $N$ are not used to avoid overfitting.}
\end{figure}

\clearpage

\begin{figure}
\centering
\includegraphics[width=\textwidth]{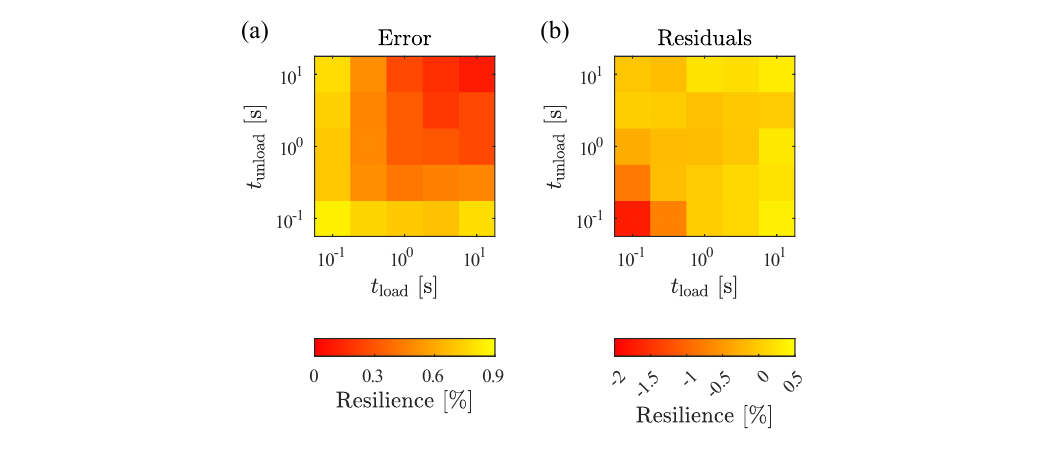}
\caption{\label{figS2}\textbf{The error of the experimental resilience values for the polyurethane elastomer and the residuals of the simulated results using the generalized Maxwell model indicate an excellent fit to the experimental data.} \textbf{(a)} The standard error of the mean of the values for resilience from the experiment (left panel of Fig.~\ref{fig2}a) is computed over 6 trials using different polyurethane elastomer samples cut from the same batch. \textbf{(b)} The residuals of the simulated resilience using the generalized Maxwell model (middle panel of Fig.~\ref{fig2}a) for the polyurethane elastomer indicate an excellent fit to the experimental data with a difference of less than 2\%. The largest difference occurs at the fast loading and unloading rates, which is likely attributed to the smoothing of the triangular wave generated by the DMA at high rates.}
\end{figure}

\clearpage

\begin{figure}
\centering
\includegraphics[width=\textwidth]{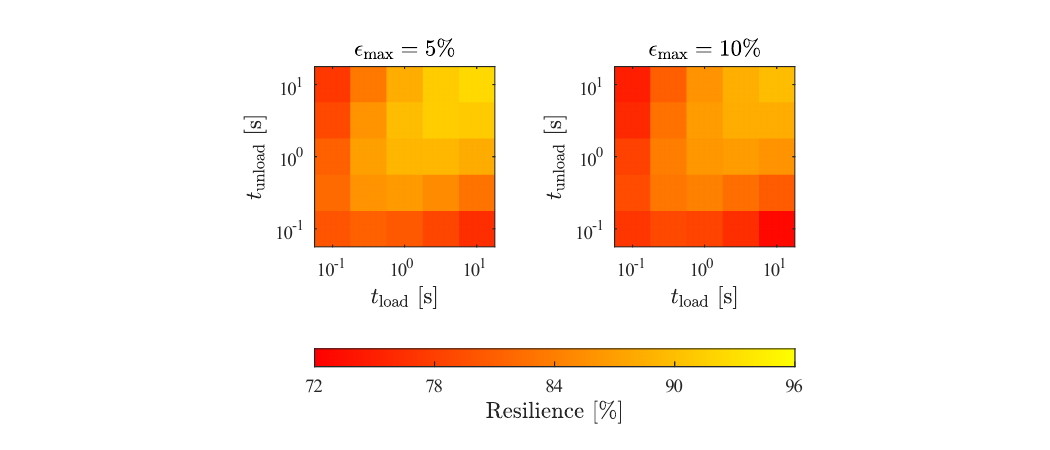}
\caption{\label{figS3}\textbf{The rate-asymmetry effect is observed at higher strains for a polyurethane elastomer.} Samples of the polyurethane elastomer are tested at symmetric and asymmetric loading and unloading rates at 5\% and 10\% maximum strain. The loading and unloading rates need to match to yield a high resilience as observed for the 1\% maximum strain case (left panel of Fig.~\ref{fig2}a). The overall resilience slightly decreases from an increase in the maximum strain and is likely caused by nonlinear effects.}
\end{figure}

\clearpage

\begin{figure}
\centering
\includegraphics[width=\textwidth]{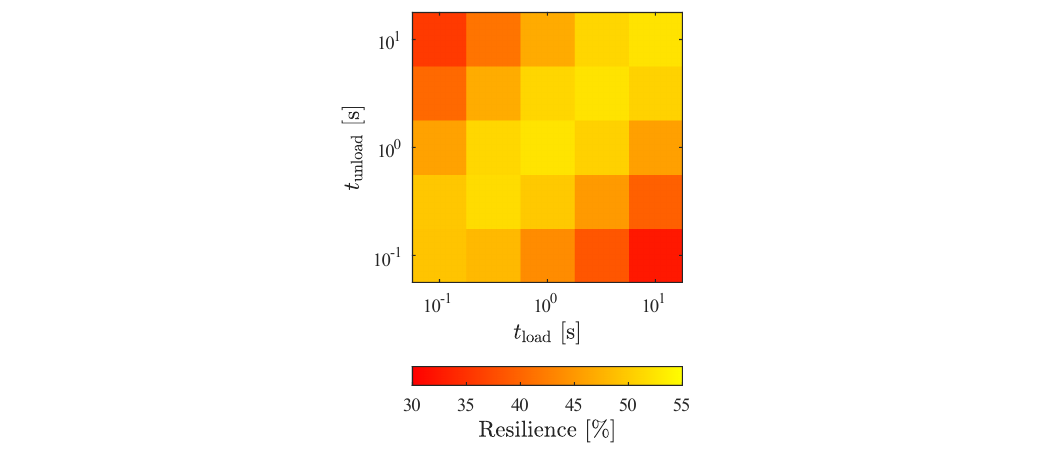}
\caption{\label{figS4}\textbf{The rate-asymmetry effect is observed for the neoprene elastomer.} The neoprene elastomer, a more dissipative material compared to the polyurethane elastomer, is tested at symmetric and asymmetric loading and unloading rates at 1\% maximum strain. The loading and unloading rates need to match to yield a high resilience as observed for the polyurethane elastomer (left panel of Fig.~\ref{fig2}a).}
\end{figure}

\clearpage

\begin{figure}
\centering
\includegraphics[width=\textwidth]{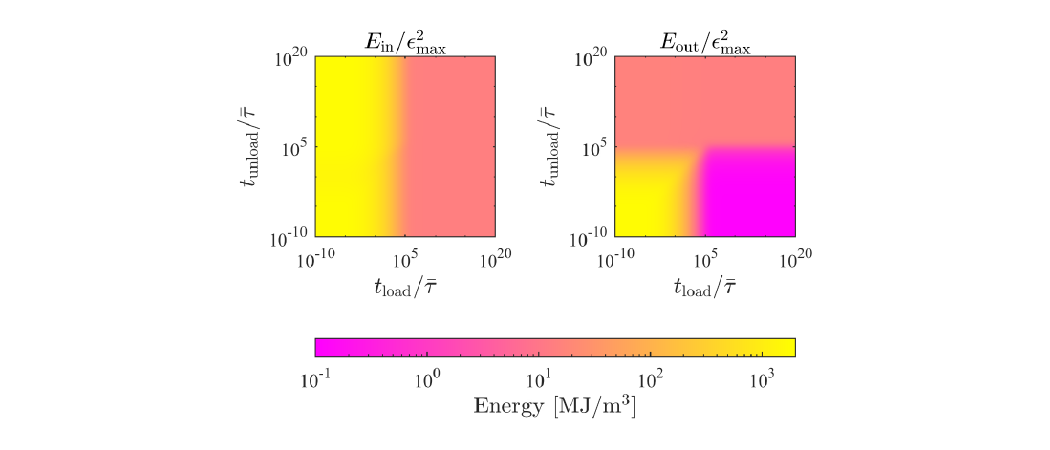}
\caption{\label{figS5}\textbf{A high resilience does not correspond to a high energy output.} The energy input and output (normalized by $\epsilon_\mathrm{max}^2$ in strain units) for the resilience profile of the polyurethane elastomer (right panel of Fig.~\ref{fig2}a) depicts a high energy output at the fast loading and unloading rate extreme ($t_\mathrm{load},t_\mathrm{unload} << \bar{\tau}$) but a low energy output at the slow loading and unloading rate extreme ($t_\mathrm{load},t_\mathrm{unload} >> \bar{\tau}$) although both extremes exhibit a high resilience (yellow squares on right panel of Fig.~\ref{fig2}a).}
\end{figure}

\clearpage

\begin{figure}
\centering
\includegraphics[width=\textwidth]{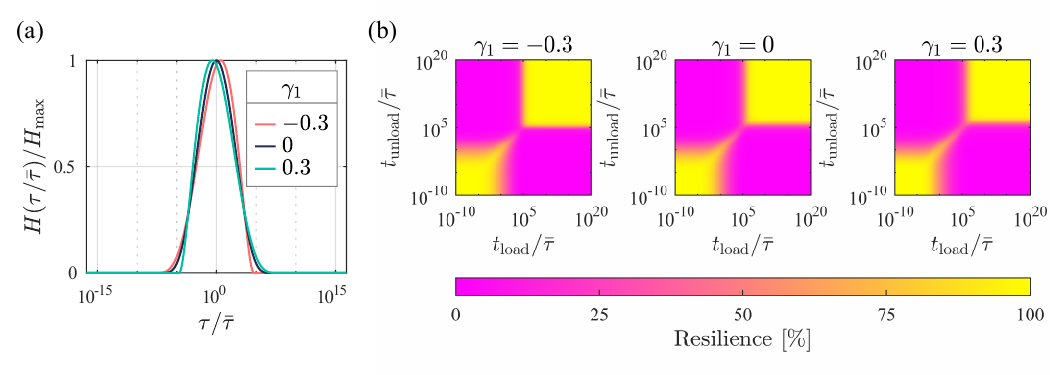}
\caption{\label{figS6}\textbf{The skewness of a relaxation spectrum does not have a pronounced impact on resilience.} \textbf{(a)} The fitted Pearson distribution parameters (blue curve of Fig.~\ref{fig2}c) for the discrete relaxation spectrum of the polyurethane elastomer is used with varied values for the skewness, $\gamma_1$, over the $\log\tau$ scale. The experimental value for skewness is $-0.0847$. \textbf{(b)} Compared to a change in $E_0/E_\infty$ (Fig.~\ref{fig3}) and $\log\sigma$ (Fig.~\ref{fig4}), a change in the skewness of the relaxation spectrum does not significantly impact the resilience profile.}
\end{figure}

\clearpage

\begin{figure}
\centering
\includegraphics[width=\textwidth]{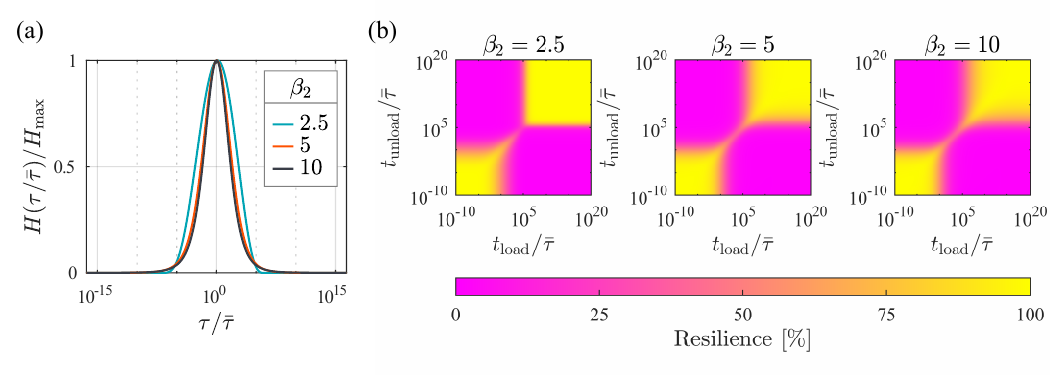}
\caption{\label{figS7}\textbf{A higher kurtosis for a relaxation spectrum broadens the resilience transition between the long timescale extreme and the low resilience regions.} \textbf{(a)} The fitted Pearson distribution parameters (blue curve of Fig.~\ref{fig2}c) for the discrete relaxation spectrum of the polyurethane elastomer is used with varied values for the kurtosis, $\beta_1$, over the $\log\tau$ scale. The experimental value for kurtosis is $2.55$ \textbf{(b)} A higher kurtosis broadens the transition between the high resilience extreme at long timescales (yellow squares on upper right of panels) and the regions of low resilience that are significantly impacted by rate-asymmetry (pink regions).}
\end{figure}

\clearpage

\begin{figure}
\centering
\includegraphics[width=\textwidth]{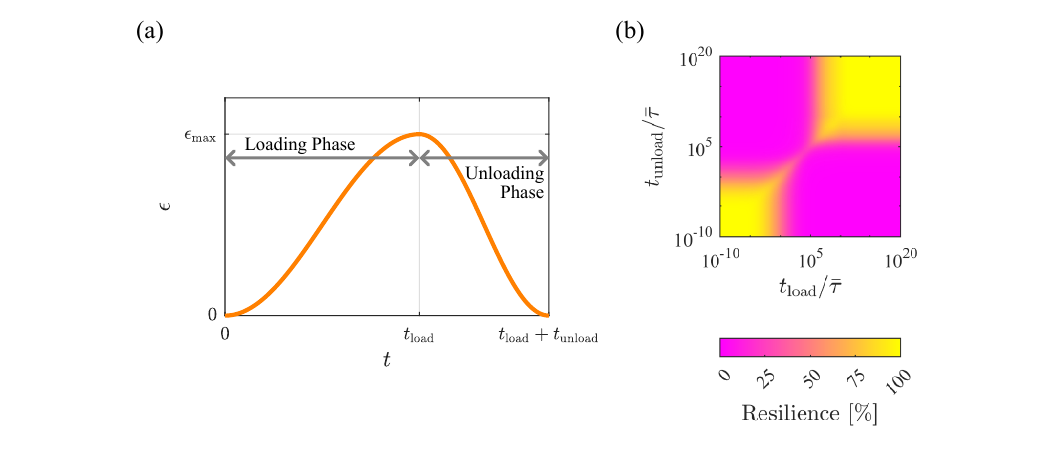}
\caption{\label{figS8}\textbf{Harmonic loading and unloading yields a similar resilience profile to that of linear loading and unloading (right panel of Fig.~\ref{fig2}a)}. \textbf{(a)} The strain versus time profile for harmonic loading and unloading consists of smooth sinusoidal functions. \textbf{(b)} Simulations were performed with harmonic loading and unloading using the discrete relaxation spectrum for the polyurethane elastomer (see Supplemental Information: Generalized Maxwell Model \& Resilience Computation: Linear Loading \& Unloading). The resilience was computed at the third cycle.}
\end{figure}

\clearpage
\renewcommand{\figurename}{Table}
\renewcommand{\thefigure}{S\arabic{figure}}

\setcounter{figure}{0}

\begin{figure}
\centering
\includegraphics[width=0.4\textwidth]{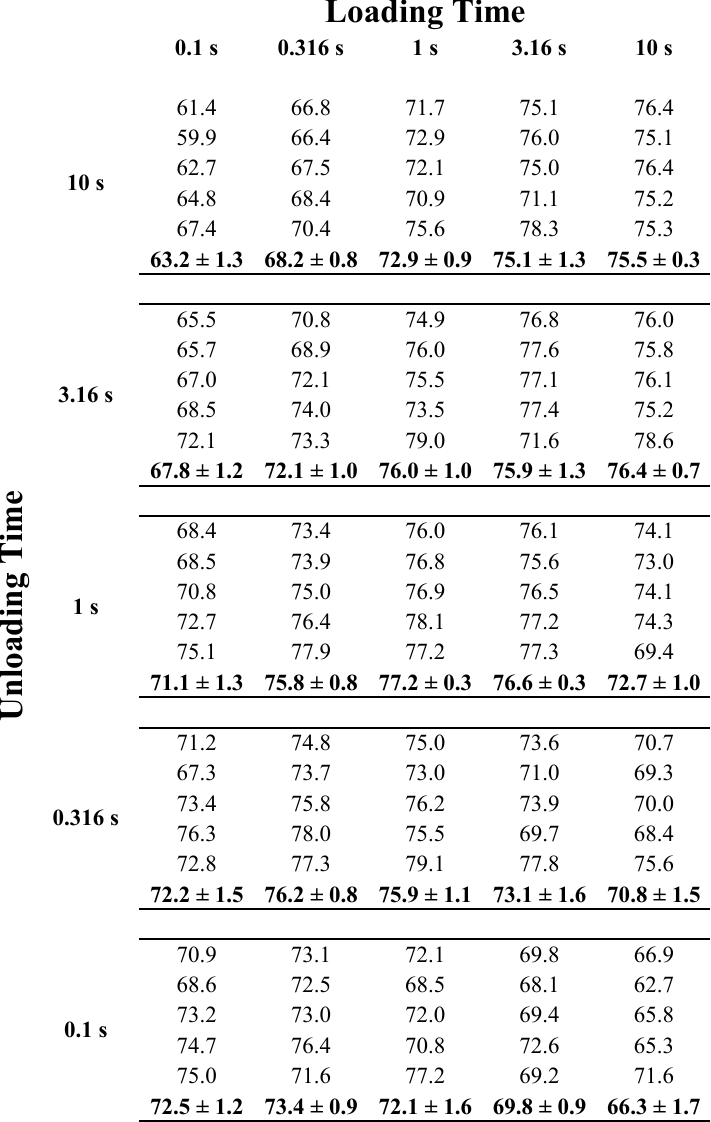}
\caption{\label{tabS1} \textbf{The full data set of the resilience of American bullfrog plantaris tendons.} Within each combination of loading/unloading times, the rows are the measured resilience for an individual tendon (with 5 tendons total measured). The bold numbers are the mean resilience at each loading/unloading combination across all 5 tendons with $\pm$ standard error of the mean resilience reported.}
\end{figure}

\clearpage

\begin{figure}
\centering
\includegraphics[width=0.4\textwidth]{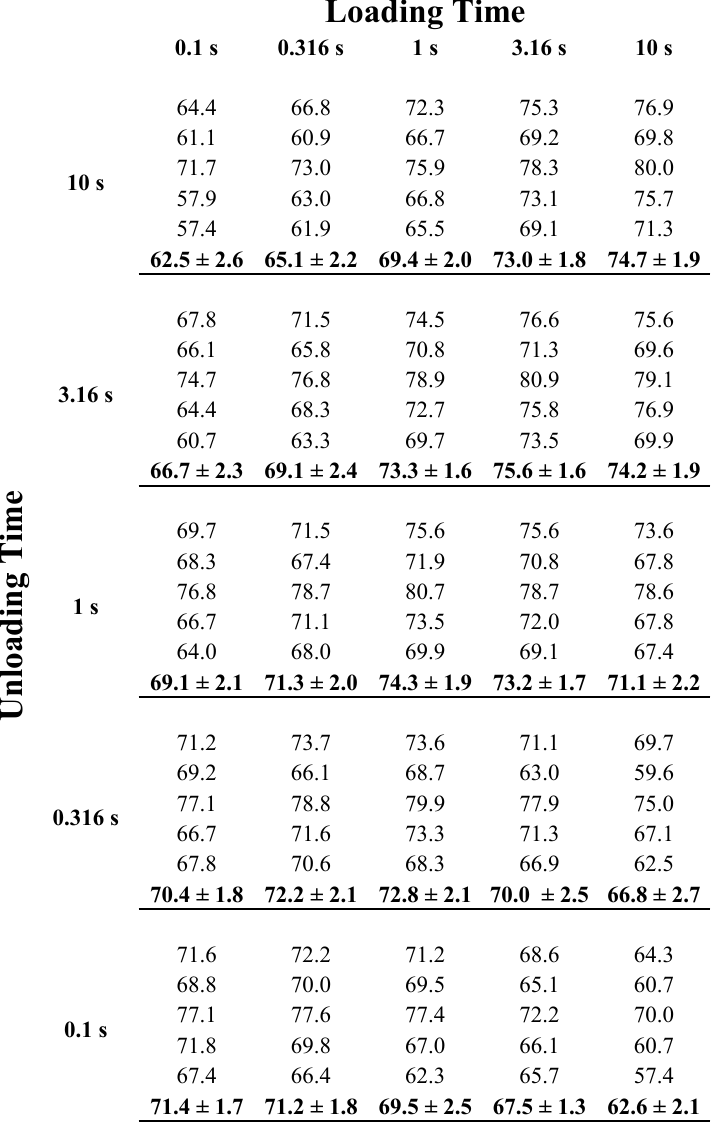}
\caption{\label{tabS2} \textbf{The full data set of the resilience of guinea fowl gastrocnemius tendons.} Within each combination of loading/unloading times, the rows are the measured resilience for an individual tendon (with 5 tendons total measured). The bold numbers are the mean resilience at each loading/unloading combination across all 5 tendons with $\pm$ standard error of the mean resilience reported.}
\end{figure}

\clearpage

\end{document}